# Effect of the chemical pressure on SDW and superconductivity in undoped and 15%F doped $La_{1-y}Y_yFeAsO$ compounds.

M.Tropeano[a,d], C.Fanciulli[a], F.Canepa[b,c], M.R.Cimberle[b], C.Ferdeghini[a], G.Lamura[a], A.Martinelli[a], M.Putti[a,d,e], M.Vignolo[a] and A.Palenzona[a,c]

[a]*CNR/INFM-LAMIA Corso Perrone 24, 16152 Genova, Italy*
[b]*CNR-IMEM, Dipartimento di Fisica, Via Dodecaneso 33, 16146 Genova, Italy*
[c]*Dipartimento di Chimica e Chimica Industriale, Via Dodecaneso 31, 16146 Genova, Italy*
[d]*Dipartimento di Fisica, Via Dodecaneso 33, 16146 Genova, Italy*
[e]*Applied Superconductivity Center, National High Magnetic Field Laboratory, Florida State University, 2031 E. Paul Dirac Dr.,Tallahassee, FL 3231, USA*

## ABSTRACT

We present a study concerning the partial substitution of yttrium at the lanthanum site of the undoped LaFeAsO and superconducting $LaFeAsO_{0.85}F_{0.15}$ compounds. We prepared samples with a nominal yttrium content up to 70% producing simultaneous shrinkage of both the *a*- and *c*-lattice parameters by 1.8% and 1.7%, respectively. The chemical pressure provided by the partial substitution with this smaller ion size causes a lowering of the spin density wave temperature in the undoped compounds, as well as an increase of the superconducting transition temperatures in the doped ones. The 15% fluorine-doped samples reach a maximum critical temperature of 40.2 K for the 50% yttrium substitution. Comparison with literature data indicates that chemical pressure cannot be the only mechanism which tunes drastically both $T_{SDW}$ and $T_c$ in 1111 compounds. Our data suggest that structural disorder induced by the partial substitution in the La site or by doping could play an important role as well.

## INTRODUCTION

Iron-arsenide based compounds (pnictides) have shown superconductivity with surprisingly high $T_c$. The first discovered superconductor has been $LaFeAsO_{1-x}F_x$ with $T_c$ as high as 26 K [1]. This compound of chemical formula $REFeAsO_xF_{1-x}$ (RE being a rare earth) belongs to the so called "1111" family. After only one month $T_c$ has doubled thanks to a substitution of the La by different rare earth (RE) elements (Sm, Ce, Nd, Pr and Gd) yielding an increase up to 52 K with Pr[2] and 55 K with Sm[3]. Almost the same high critical temperature has been reported in hole doped oxygen deficient $REFeAsO_{1-\delta}$ compounds with a maximum critical temperature of 51 K, 54 K and 55 K in Pr, Nd and Sm systems respectively[4, 5]. Nowadays $T_c$ seems to have reached its maximum value, (56 K) in the electron doped $Gd_{1-x}Th_xFeAsO$[6] .
The La system presents the lowest $T_c$ within the 1111 family, nearly one half of the maximum value, likely due to the large ionic radius of La. This aspect makes La-1111 one of the best systems to investigate the effects of the external and chemical pressure on $T_c$. By applying an external pressure of 4 GPa to optimally doped $LaFeAsO_{1-x}F_x$, $T_c$ has been raised up to 43K, then slowly dropped off to 9 K for larger pressures [7]. F-doped and O-deficient La1111 under external pressure has shown how the increase in $T_c$ is related to doping: $LaFeAsO_{1-\delta}$ under 1.5 GPa[8] has shown a superconducting $T_c$ onset of about 50 K, which is the highest record in La-based system.
Also the role of chemical pressure has been explored by partial substitution of the RE element in the $La_{1-y}RE_yFeAsO_{1-\delta}$ compound: $T_c$ up to 43 K has been obtained in the 40% yttrium substituted compound[9]. In the case of RE=Sm a monotonic rise of $T_c$ from 30 K (y=0) to (y=1) 55 K[10] has been reported.
All these experimental results clearly evidence that both the substitution with smaller ions and the application of an external pressure cause lattice shrinkage, even if the mechanisms involved are different: the chemical pressure is isotropic whereas external pressure may induce an anisotropic shrinkage because of the pnictide layered structure. In both cases a strong correlation between lattice shrinkage and $T_c$ can be established.
The external pressure produces significant effects also on the structural transition and/or Spin Density Wave (SDW) ordering related temperatures of the parent compound. In a weakly F doped SmFeAsO sample a reduction of $T_{SDW}$ with pressure has been observed[11]. A suppression of the SDW with pressure has been put in evidence also in $CaFe_2As_2$[12] and in $K_xSr_{1-x}Fe_2As_2$[13] (122 family). In ref. 12 it has been argued that the simultaneous increase of $T_c$ and decrease of $T_{SDW}$ with pressure are evidences that pressure induces doping through charge transfer from the charge reservoir block. These aspects have never been considered in systems where the chemical pressure is varied.
In this report we investigate the effect of chemical pressure on $T_c$ and $T_{SDW}$ through the partial substitution of the La ions by Y in the pure and 15% F-doped LaFeAsO compounds. Increasing the Y content the cell parameters reduce monotonically. In the undoped compounds this shrinking results in a monotone decrease of $T_{SDW}$ down to 120 K, while in the 15% F-doped system the critical temperature increases reaching a maximum of about 40 K for 50% yttrium substitution and then decreases with further increasing yttrium content.

## EXPERIMENTAL

### A. Synthesis
The samples of the series $La_{1-y}Y_yFeAsO_{1-x}F_x$, (y= 0, 0.30, 0.50 and 0.70 and x= 0 and 0.15) were synthesized at *ambient pressure* by a solid state reaction method using high purity (La, Y)As, $Fe_2O_3$, Fe and $FeF_2$ with a two steps reaction procedure[14,15]. A stoichiometric pellet was pressed and reacted in a welded tantalum crucible closed in quartz ampoules at 1000°C for 70 hours. The pellet was subsequently grinded, mixed, pressed in a new pellet an annealed with the same procedure at 1250°C for 25 hours. All the handling and manipulation of the samples was carried out in a glove box where the working atmosphere was continuously purified to less than 1 ppm $H_2O/O_2$. The use of tantalum crucibles prevented losses of fluorine, the dopant responsible for the superconductivity, even if some consumption of arsenic by tantalum was observed. The final product was a black and hard cylinder, with a density of about 80 % of the theoretical value, which has been cut for measurements. The same procedure was also used without success to synthesize the compounds with y=1, YFeAsO and $YFeAsO_{0.85}F_{0.15}$: only a mixture constituted by $Y_2O_3$, YAs, FeAs, $Fe_2As$ (and YOF in the F-doped sample) was obtained. Probably, in this case, an high pressure synthesizing method followed by a special rapid quench process would be necessary, as reported by Jie Yang and co-workers[16].

### B1. Sample characterization: structural analysis
Phase identification was performed by X-ray powder diffraction (XRPD; PHILIPS PW3020; Bragg-Brentano geometry; $CuK_\alpha$; range 15 – 120° $2\theta$ ; step 0.020°; sampling time 10 s); the crystal structures of the samples were refined in the space group *P*4/*nmm* - 129 (origin choice 2) according to the Rietveld method using the FullProf software; by means of a $LaB_6$ standard an instrumental resolution file was obtained and applied during refinements in order to detect possible micro-structural contribution to XRPD peak shape.
Figure 1 shows the Rietveld refinement plot obtained for the sample with nominal composition $(La_{0.70}Y_{0.30})FeAs(O_{0.85}F_{0.15})$, containing (La,Y)As and (La,Y)OF as secondary phases, whose amounts result amounts result 6.4% and 8%, respectively.

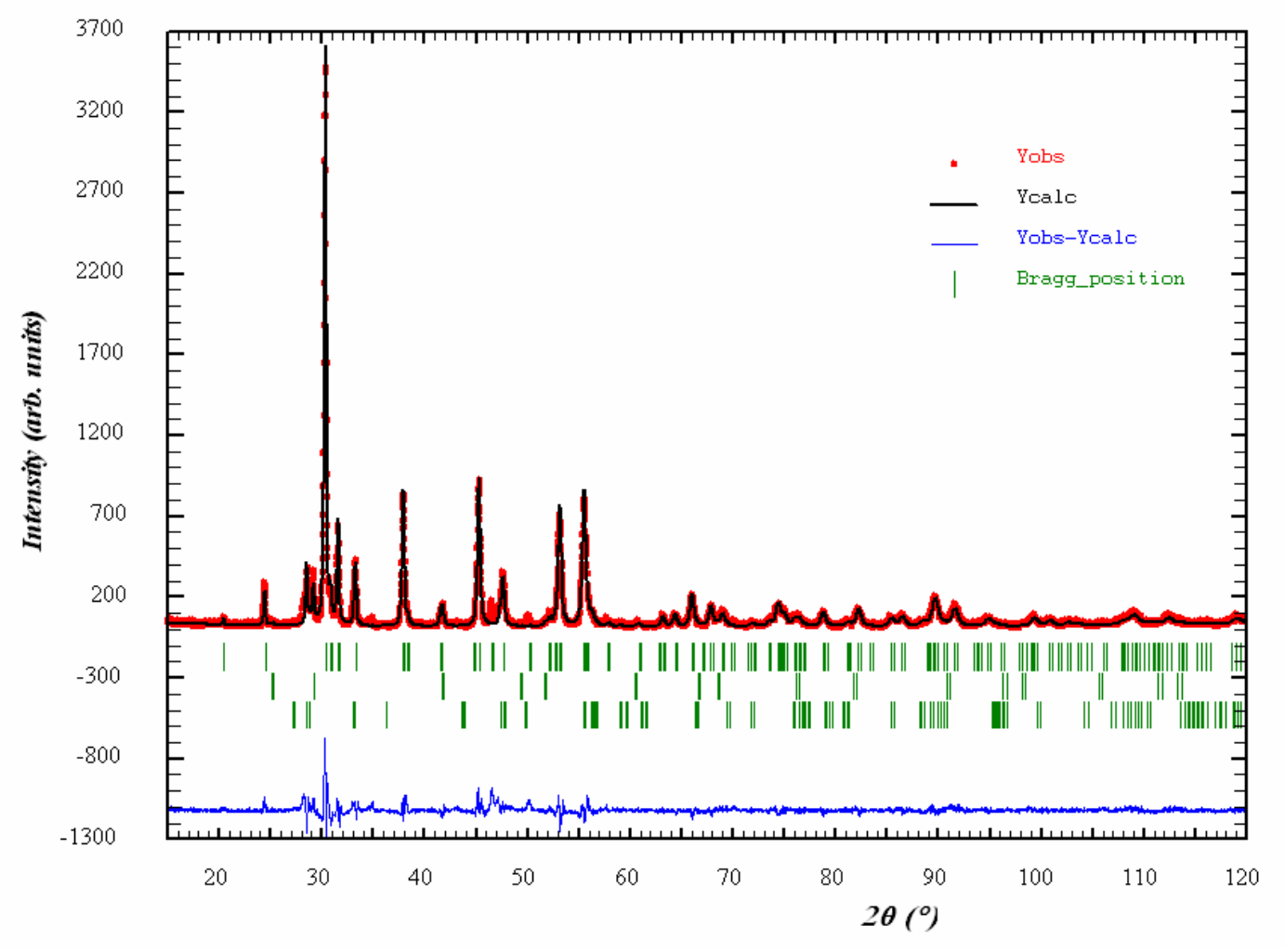

**Figure 1.** (Color online) Rietveld refinement plot obtained for the samples with nominal composition $(La_{0.70}Y_{0.30})FeAs(O_{0.85}F_{0.15})$.

The effectiveness of the substitution is proved by the monotonic decrease of the *a*- and *c*-lattice parameters obtained from Rietveld refinement with Y content as shown in Figure 2. Information on the chemical homogeneity of the samples was obtained analyzing the broadening XRPD lines by means of the Williamson-Hall plot method [17]. Generally, in case size effect are negligible, a straight line passing through the origin have to be observed, whereas the slope provides the lattice strain. When broadening is not isotropic, size and strain effects along some crystallographic directions can be obtained by considering different orders of the same reflection. For solid solutions, the $\Delta d/d$ (*d*: interplanar spacing) term can have also a contribution from chemical fluctuations and inhomogeneities inside the sample, because the larger the distribution of lattice parameters the broader the peaks. Broadening related to chemical inhomogeneity is generally anisotropic, depending on the symmetry of the crystal system[18]. Williamson-Hall plots for all the F-doped samples (not showed) evidence two main features 1) size contribution is negligible since a straight line passing through the origin can be traced and 2) broadening related to microstrain and/or chemical fluctuations is isotropic, indicating a good chemical homogeneities of our samples since they crystallize in the tetragonal system.

Figure 2 shows the cell parameters of $La_{1-y}Y_yFeAsO$ and $La_{1-y}Y_yFeAsO_{0.85}F_{0.15}$ (square symbol) as a function of yttrium nominal content. As expected, in the undoped compounds, starting from the y=0 sample (*a*= 4.033 Å, *c*= 8.744 Å) the effect of the substitution results in a monotonic decrease by 1.8% and 1.7% of *a*- and *c*-lattice parameters respectively for the highest yttrium content (*a*= 3.962 Å, *c*= 8.597 Å). The 15% fluorine doping shows a further shrinkage of about 0.3% of these cell parameters, indicating the covalent character of the intra-layer chemical bonding due to the smaller radius of fluorine as compared to oxygen one[3].

Substitution at the La site has been earlier reported in oxygen deficient $La_{1-x}Y_xFeAsO_{1-\delta}$ compounds produced by high pressure synthesis process[9]: the *a*- and *c*-lattice parameters exhibit a weak decrease, 0.92% and 0.88% respectively for 40% yttrium substitution. With the y=0.7 sample we were able to shrink the axes by more than twice, which allows us to investigate the effect of chemical pressure in a more extended range.

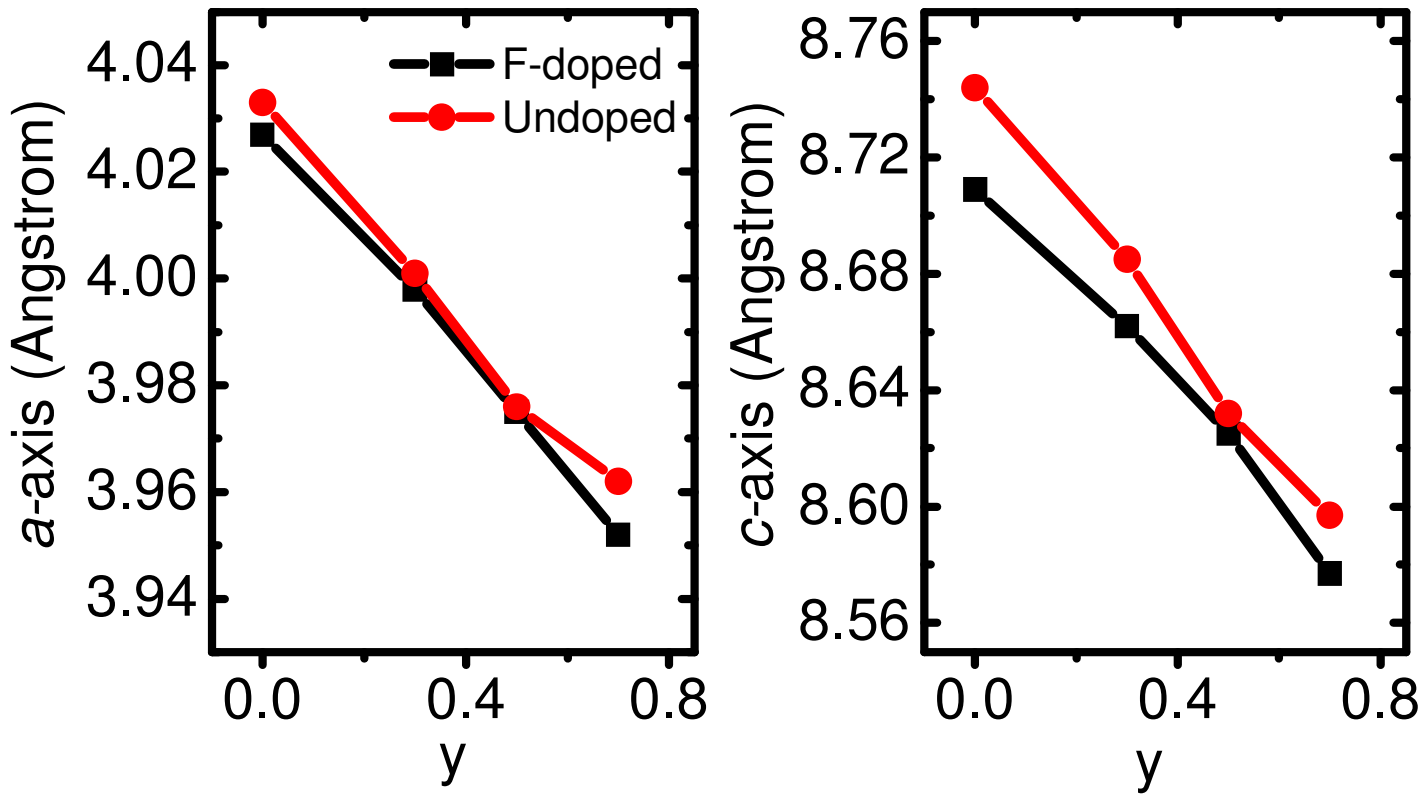

**Figure 2.** (Color online) Cell parameters of $La_{1-y}Y_yFeAsO$ (circle symbol) and $La_{1-y}Y_yFeAsO_{0.85}F_{0.15}$ (square symbol) as a function of yttrium nominal content.

## B2. Sample characterization: resistive measurements

Resistive measurements were taken using a standard four probe technique. In Figure 3 resistivity versus temperature measurements of the undoped compounds are shown: data have been normalized to the room temperature values and shifted for a better visualization.

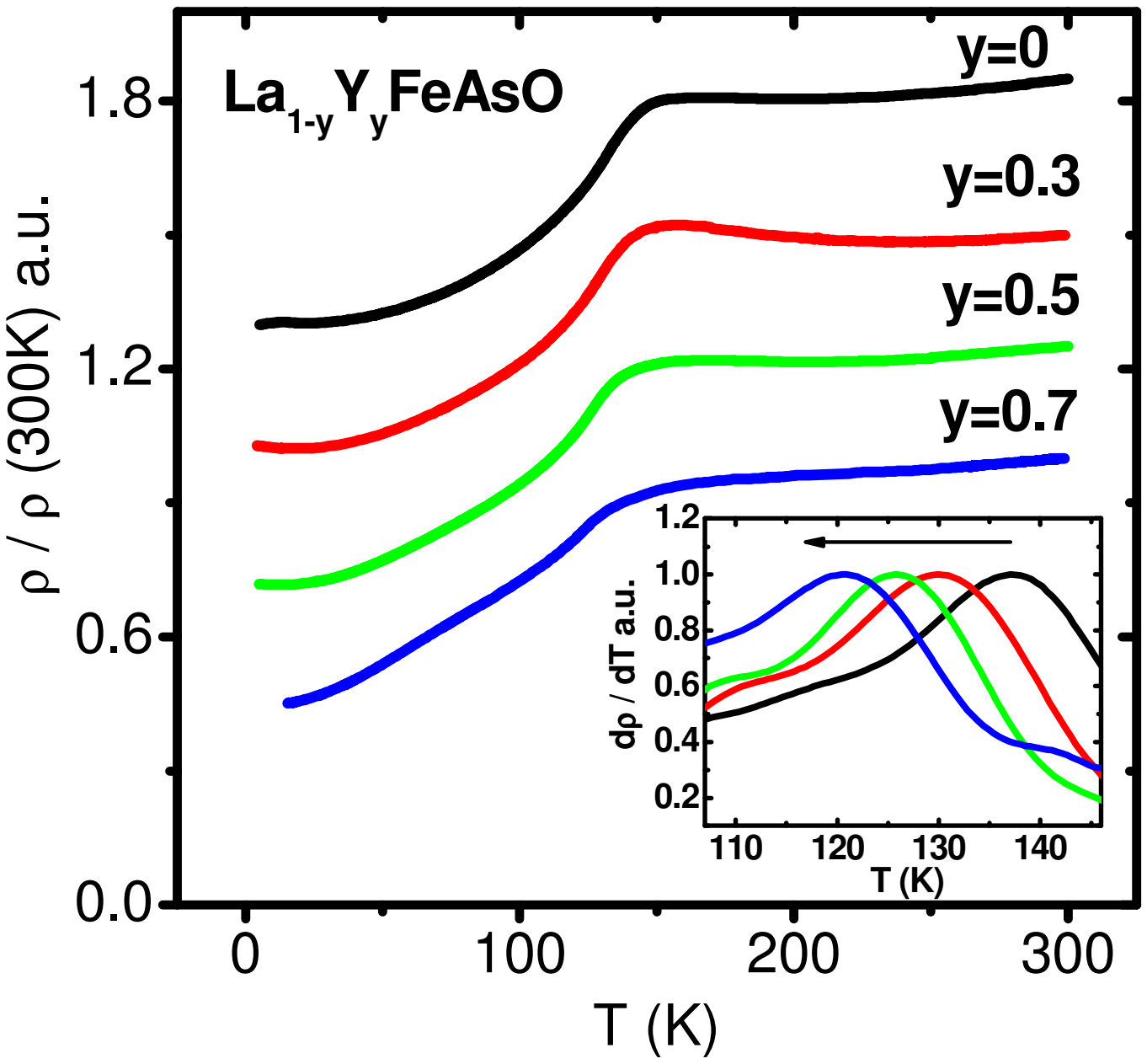

**Figure 3.** (Color online) $La_{1-y}Y_yFeAsO$ normalized resistivity versus temperature behavior. The inset shows the shift toward low temperatures of the maximum of normalized dρ / dT with increasing the yttrium content.

Upon cooling, ρ(T) of undoped samples presents typical features of the iron based oxypnictides compounds: a maximum, followed by a sharp drop with an inflection point at $T_{drop}$ defined as max of the first derivative dρ/dT. We recall that in LaFeAsO[19, 20] and in PrFeAsO compounds[21] the maximum of ρ has been attributed to the occurrence of the structural transition while the maximum of the first derivative to the SDW transition. For the $SmFeAsO_{1-x}F_x$ system the two transitions seem to coincide [22].

By increasing the Yttrium content a broadening of the maximum in the ρ(T) is observed up to its disappearing in the case of 70% Y substituted sample. Furthermore $T_{drop}$ is shifted to lower temperatures starting from a value of $T_{drop}$ = 138 K in the LaFeAsO down to 120 K in the 70% Y substituted compound. In the following we assume $T_{drop} \equiv T_{SDW}$.

In Figure 4 ρ(T)/ ρ(300K) for 15% F-doped compounds are plotted and shifted for better visualization.

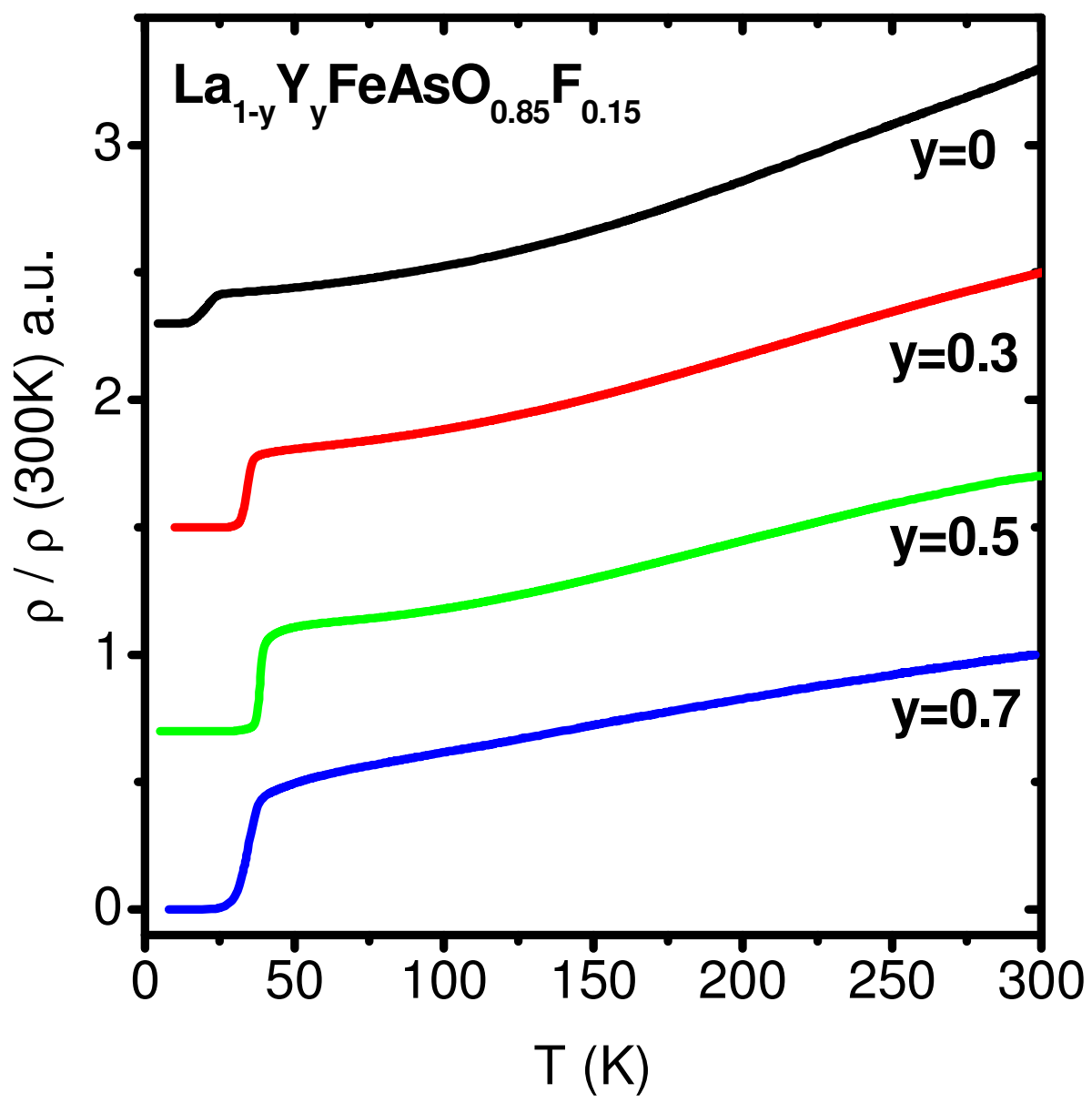

**Figure 4.** (Color online) Temperature dependence of the resistivity for $La_{1-y}Y_yFeAsO_{0.85}F_{0.15}$ compounds.

The critical temperatures $T_c$ are evaluated by considering the intersection of the two straight lines drawn on ρ(T) data in the normal state just above $T_c$ and its steepest part in the SC state, as reported in Ref. 7. $T_c$ increases with the Y content, reaches a maximum $T_c$ = 40.2K for y=0.5 and then decreases. All data for undoped and F-doped samples are collected in Figure 5, which shows the evolution of $T_{SDW}$ and $T_c$ as a function of the yttrium content.

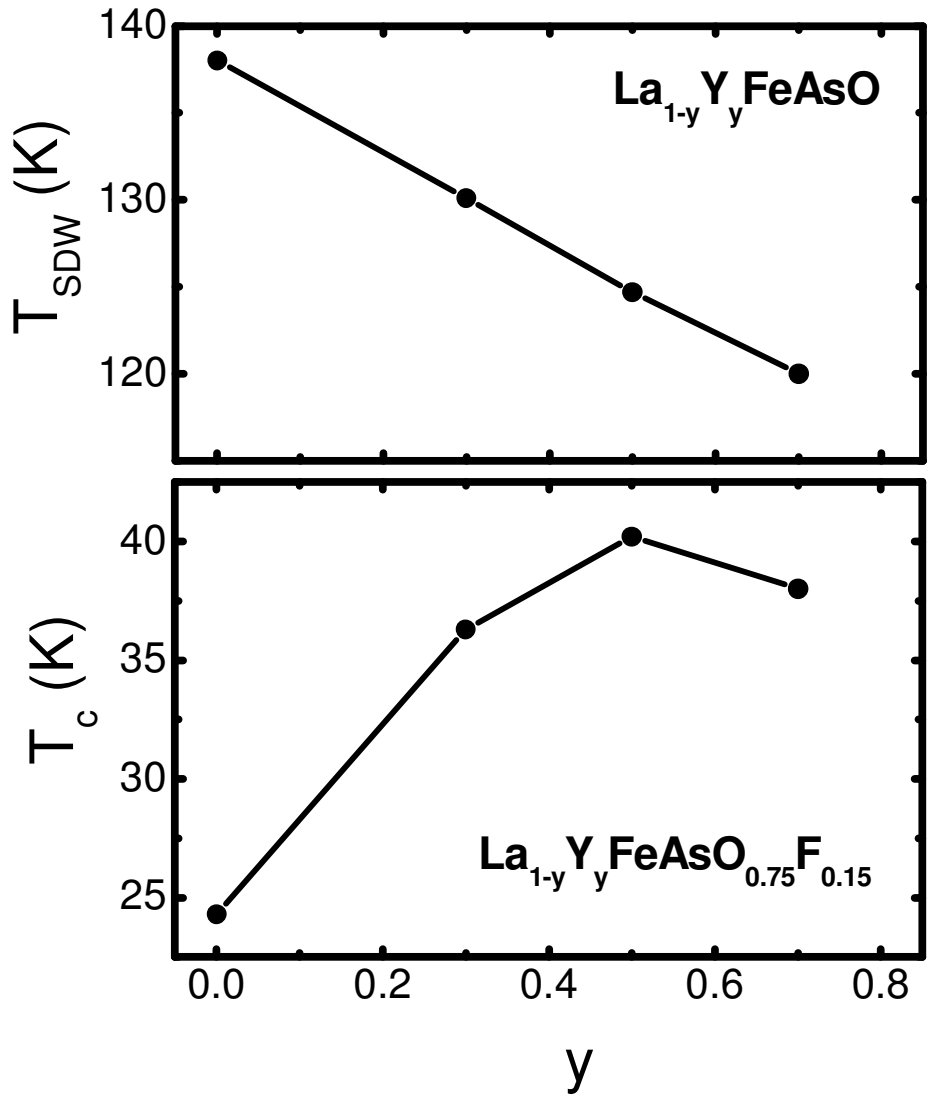

**Figure 5.** Temperature evolution of $T_{SDW}$ and $T_C$ as a function of yttrium content for $La_{1-y}Y_yFeAsO$ and $La_{1-y}Y_yFeAsO_{0.85}F_{0.15}$ compounds.

## B3. Sample characterization: DC magnetic measurements

DC magnetization was performed in a DC-SQUID magnetometer (MPMS, Quantum Design). The temperature dependence of the Zero Field Cooled (ZFC) and Field Cooled (FC) dc-magnetization (*m*) taken with an applied dc field $\mu_0H=10^{-3}T$ for the $La_{1-y}Y_yFeAsO_{0.85}F_{0.15}$ compounds are plotted in Figure 6. The drop in *m*(T) indicates the onset of the superconductivity.

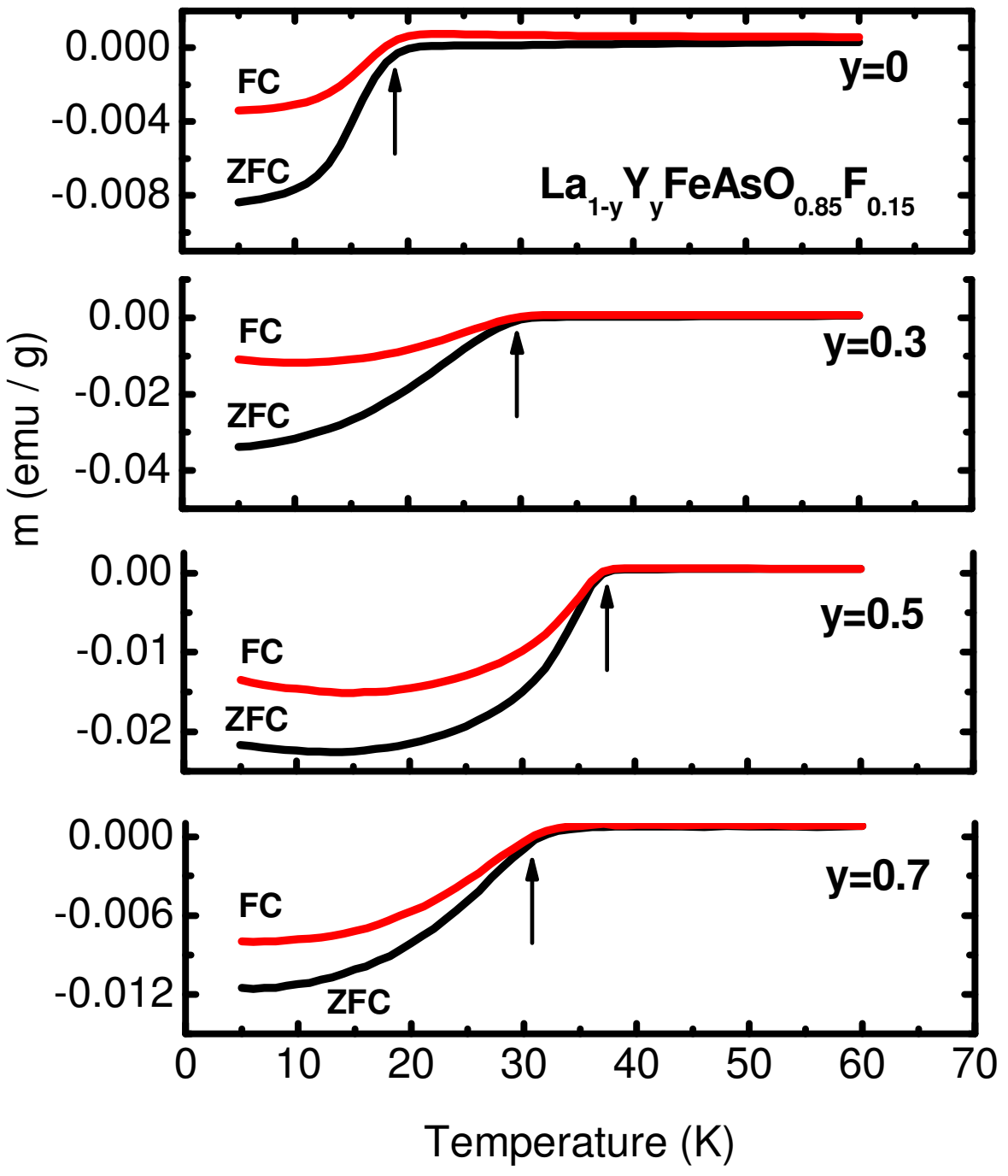

**Figure 6.** (Color online) ZFC and FC magnetization for $La_{1-y}Y_yFeAsO_{0.85}F_{0.15}$ compounds. The black arrows indicate the critical temperatures $T_c$ evaluated in the same way as the resistivity ones, that is through the intersection of the lines corresponding to the normal state above $T_c$ and the steepest part of the magnetization in the superconducting one.

The shielding values is about -1 for the samples with y=0, whereas it becomes rather low with increasing Yttrium content: 30% for the y=0.3 sample and down to 15% for the y=0.7 sample. The low shielding values are indicative of bad weak-links between grains and of the effect of the penetration depth $\lambda$ ( ~300nm at low temperature[23]) on the shielded grain volume (the mean grain size is about few μm). The presence of spurious magnetic phases may be inferred also from the positive offset shown by *m*(T) above the SC transition and from the slight upturn of magnetization at low temperature observed in some curves. $T_c$ values evaluated from magnetization measurements, indicated by the black arrows in figure 5, are a bit lower than those evaluated by the resistivity, but present the same behavior as a function of y.

## DISCUSSION

We have shown that both $T_c$ in $La_{1-y}Y_yFeAsO_{0.85}F_{0.15}$ samples and $T_{SDW}$ in the parent $La_{1-y}Y_yFeAsO$ samples present a clear dependence on the Y content. Since the lattice parameters monotonically shrink with increasing Y content, the evolution of $T_c$ and $T_{SDW}$ can be discussed as a function of chemical pressure.

Figure 7 (a) shows the critical temperatures of the $La_{1-y}Y_yFeAsO_{0.85}F_{0.15}$ series as a function of *a*-axis with data collected from literature. As it was previously noted for oxygen deficient compounds[9], the literature data of the 1111 family seems to stay into a common line which describes the increase of $T_c$ with the decrease of the lattice parameters, showing a clear indication in favor of the role played by the chemical pressure in increasing $T_c$.

Our 15%F doped La-Y series, instead, by increasing the chemical pressure, show a peculiar dome-like behavior, compressed and shifted in the right-bottom side of the graph in respect to that of the 1111 family with high-$T_c$ rare earth: with decreasing the *a*-axis, $T_c$ increases showing a maximum which is about 40 K for a=3.975 nm, and then decreases for smaller *a*-axis.

In Figure 7 (b) we plot $T_{SDW}$ of the $La_{1-y}Y_yFeAsO$ series as a function of *a*-lattice parameter with data collected from literature on the 1111 family, with RE=Ce, Pr, Nd, Sm, Gd, Tb. In all these data $T_{SDW}$ was estimated by the maximum of $d\rho/dT$. The graph shows two distinct curves which monotonically rise with increasing the a-axis. The upper line contains all the high-$T_c$ rare earths. The lower line contains the La-Y series. This behavior resembles the one reported for $T_{SDW}$ versus external pressure in the 1111[11] and 122[12,13] families, confirming the effectiveness of the Y substitution in La site in producing chemical pressure.

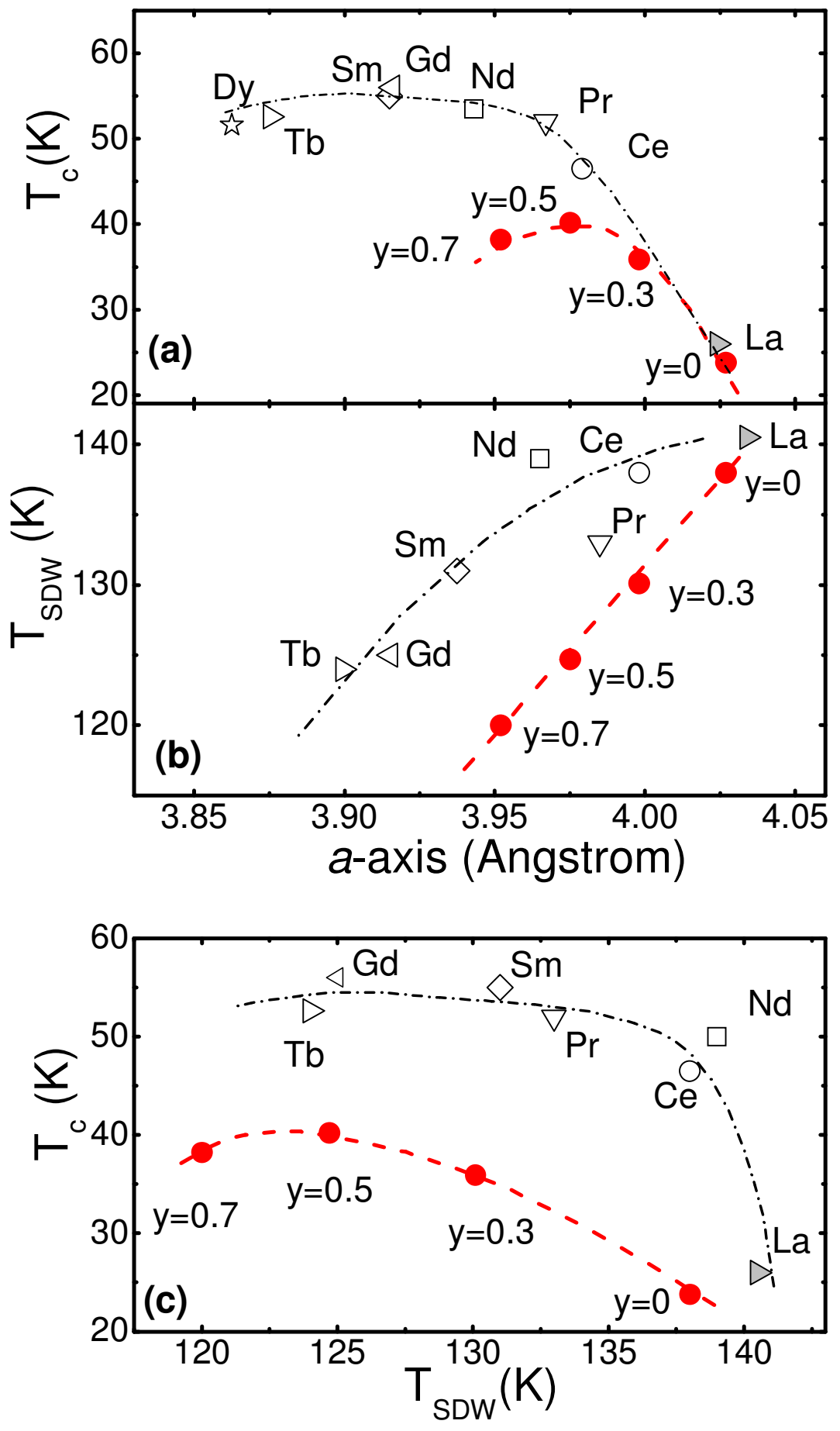

**Figure 7.** (Color online) (a) $T_c$ and (b) $T_{SDW}$ vs the *a*-axis, for the $La_{1-y}Y_yFeAsO_{0.75}F_{0.15}$ samples (full circle) in comparison with data collected from literature for 1111 compounds with different rare earths. (c) $T_c$ versus $T_{SDW}$ for La-Y series and other rare earths. The SC samples used for comparison are: $LaFeAsO_{0.9}F_{0.1-\delta}$ (Ref. 24), $CeFeAsO_{0.84}F_{0.16}$ (Ref. 25), $PrFeAsO_{0.84}F_{0.16}$ (Ref. 2), $NdFeAsO_{1-\delta}$ (Ref.5), $SmFeAsO_{0.9}F_{0.1}$ (Ref. 3) 3, $Gd_{0.8}Th_{0.2}FeAsO$ (Ref. 6), $TbFeAsO_{1-\delta}$ (Ref. 26), $DyFeAsO_{1-\delta}$ (Ref.26). $T_{SDW}$ data for undoped REFeAsO are, when different data are present for the same compound, the average between the maximum and minimum value reported: La ($T_{SDW}$= 138 K[19] and $T_{SDW}$=143 K[27] ); Ce ($T_{SDW}$= 136 K[27]and $T_{SDW}$= 140 K[28]); Pr ($T_{SDW}$= 127 K[21] and $T_{SDW}$= 139 K[27]); Nd ($T_{SDW}$= 137 K[29] and $T_{SDW}$ = 141 K[27]); Sm ($T_{SDW}$= 131 K[26]); Gd ($T_{SDW}$= 125 K[6]) and Tb ($T_{SDW}$= 124 K[26]).

The twofold effects of increasing $T_c$ and decreasing $T_{SDW}$ was considered as an evidence of doping induced by pressure[13]. Our data cannot exclude that chemical pressure can contribute to doping, but the indication we can drawn is that this is a small effect. In fact, in the undoped compound, starting from y=0, the a-axis is shrunk by 1.8 % without observing the occurrence of superconductivity, while the superconductivity occurs after F substitution which causes a much lower further shrinkage (about 0.3%). Therefore we conclude that in the 1111 family, doping with F, as well as the O vacancies, is able to suppress the SDW and favor superconductivity much more effectively than chemical pressure does.

Finally figure 7(c) shows $T_c$ as a function of $T_{SDW}$. Also in this case a correlation can be found provided that we consider separately the La-Y series and the high-$T_c$ rare earth series. The graph suggests that smaller $T_{SDW}$ favor higher $T_c$, even if a saturation around 55 K and 40 K, in the high-$T_c$ rare earth and in La-Y series, respectively, occurs. This point needs more investigation.

It is interesting to discuss the differences between La and other rare earths 1111 compounds that emerges from the discussion above. It is evident that the larger ionic radius of La cannot explain these differences because by substituting a certain amount of Y we can obtain a cell volume smaller than in the case of Ce and Pr, and comparable with that of Nd, but $T_c$ values remain smaller.
We should point out that the substitution of La with Y decreases the average lattice parameters, but, locally, this could be not equivalent to the effect exerted by ions with smaller radius. Structural disorder induced by substitution could also be detrimental to superconductivity. Moreover, an important role can be also played by different kind of doping: by substituting up to 50% of F $T_c$ does not reach 40 K , while reducing O, $T_c$ over 40 K can be obtained.[8]
To better clarify the different effects of doping and chemical pressure further studies on La-Y series with different type of doping can be useful.

## CONCLUSIONS

In order to investigate the effect of chemical pressure on normal and superconducting properties of the 1111 compounds, we have synthesized $La_{1-y}Y_yFeAsO_{1-x}F_x$ compound as a function of yttrium content and x=0 and 0.15 at ambient pressure. The progressive inclusion of Y is proved by the monotonic decreasing of the lattice parameters as a function of y content. In the undoped compounds the SDW ordering progressively shifts to lower temperature, and superconductivity does not occur. In the 15% doped samples $T_c$ progressively grows from 24 K to 40 K with increasing y from 0 to 0.5 and then decreases. Similar evolution of $T_{SDW}$ and $T_c$ have bee observed as a function of external pressure, indicating that chemical and external pressure play a very similar role.
To emphasize the role of chemical pressure we have plotted $T_{SDW}$ and $T_c$ as a function of a-axis of La-Y series in comparison with other rare earths. Both $T_{SDW}$ and $T_c$ of La-Y system evolve in similar, but distinct way respect to other high-$T_c$ rare earths-1111 compounds.
Lattices parameters have been widely varied in the La-Y series obtaining a-axis values in between those of Nd and Sm. This indicates that chemical pressure cannot be the only mechanism which tune drastically both $T_{SDW}$ and $T_c$ in 1111 compounds. Our data suggest that structural disorder induced by the partial substitution in the La site or by doping could play an important role as well.

### Acknowledgements

The authors would like to thank Chiara Tarantini (NHMFL, Tallahassee FL, USA) and Ilaria Pallecchi (CNR/INFM-LAMIA, Genova, Italy) for useful discussion. This work is partially supported by Compagnia di S. Paolo and by the Italian Foreign Affairs Ministry (MAE) - General Direction for the Cultural Promotion.